\documentclass[useAMS,usenatbib]{mn2e}
\pdfoutput=1
\usepackage{amsfonts,amsmath,amssymb}
\usepackage{times}
\usepackage{paralist}
\usepackage{graphicx}
\usepackage[normalem]{ulem}
\usepackage[colorlinks,linkcolor=blue,citecolor=blue,urlcolor=blue,pdftitle={First measurement of gravitational lensing by cosmic voids in SDSS},pdfauthor={Peter Melchior, P. M. Sutter, Erin S. Sheldon, Elisabeth Krause, Benjamin D. Wandelt}]{hyperref}
\newcommand{\hmpc}{$h^{-1}$Mpc}

\title{First measurement of gravitational lensing by cosmic voids in SDSS}
\author[Melchior et al.]{Peter Melchior$^{1,2}$\thanks{E-mail: melchior.12@osu.edu}, 
P. M. Sutter$^{1,3,4,5}$, 
Erin S. Sheldon$^6$, 
Elisabeth Krause$^{7}$,
\newauthor
Benjamin D. Wandelt$^{3,4,5,8}$\\
\\
$^{1}$Center for Cosmology and Astro-Particle Physics, The Ohio State University, Columbus, OH 43210, USA\\
$^{2}$Department of Physics, The Ohio State University, Columbus, OH 43210, USA \\
$^{3}$ UPMC Univ Paris 06, UMR7095, Institut d'Astrophysique de Paris, F-75014, Paris, France \\
$^{4}$ CNRS, UMR7095, Institut d'Astrophysique de Paris, F-75014, Paris, France \\
$^{5}$ Department of Physics, University of Illinois at Urbana-Champaign, Urbana, IL 61801, USA\\
$^{6}$ Brookhaven National Laboratory, Bldg 510, Upton, NY 11973, USA\\
$^{7}$ Department of Physics and Astronomy, University of Pennsylvania, Philadelphia, PA 19104, USA\\
$^{8}$ Department of Astronomy, University of Illinois at Urbana-Champaign, Urbana, IL 61801, USA\\
}

\begin{document}
\date{}
\pagerange{\pageref{firstpage}--\pageref{lastpage}} \pubyear{2014}
\maketitle
\label{firstpage}

\begin{abstract}
We report the first measurement of the diminutive
lensing signal arising from matter underdensities associated with cosmic
voids. While undetectable individually, by stacking the weak
gravitational shear estimates around 901 voids detected in SDSS DR7 by \citet{Sutter2012a}, we find substantial evidence for a depression of the
lensing signal compared to the cosmic mean. This depression is most
pronounced at the void radius, in agreement with analytical models of
void matter profiles. Even with the largest void sample and imaging survey available today, we cannot put useful constraints on the radial dark-matter void profile. We invite independent investigations of our findings by releasing data and analysis code to the public at \href{https://github.com/pmelchior/void-lensing}{https://github.com/pmelchior/void-lensing}.
\end{abstract}


\section{Introduction}
Voids are low-density environments, interesting both as probes of cosmology 
via their shape and size distributions~\citep[e.g.,][]{Bos2012, Sutter2012b} as well as laboratories for studying galaxy formation and modified gravity via their internal 
structure~\citep[e.g.,][]{Goldberg2004, Platen2008, Li2012, Spolyar2013}. While the existence of voids has been known since 
the earliest galaxy redshift surveys~\citep{Gregory1978,Kirshner1981}, 
it is only recently with the 
advent of high-density large-volume spectroscopic surveys such as the 
Sloan Digital Sky Survey~\citep[SDSS;][]{Abazajian2009} 
that reliable void catalogs 
have become available~\citep{Pan2011, Sutter2012a}. 

Despite the large numbers of voids detected in spectroscopic surveys, 
our knowledge is limited by the fact that observed voids are defined
by biased, sparse tracers of the underlying matter density, namely 
galaxies. On the other hand, our understanding of the evolution and 
characteristics of voids comes from analytical estimation~\citep{Furlanetto2006,Tinker2009} and 
dark matter $N$-body simulation~\citep[e.g.,][]{Colberg2005, Kreckel2011, LavauxGuilhem2011}, which require semi-analytic 
modeling \citep{DeLucia2009} or halo occupation distributions \citep{Berlind2002, Mandelbaum06.2, Zehavi2011} to reproduce the statistics of observed galaxy populations. 
Thus, the link between our theoretical 
understanding of dark matter underdensities and the voids observed 
in redshift surveys is tenuous at best.
One way to strengthen this link is to measure the matter underdensities directly by means of weak gravitational lensing. While the shear signal from all but the largest single voids will be undetectably small \citep{Amendola1999}, theoretical calculations by \citet{Krause2013} and numerical simulations by \citet{Higuchi13.1} suggest that stacking many voids will not only enable the detection of the effect with high significance, it will also constrain the radial profile of voids. By measuring the radial shear profile around a sufficient number of voids, we can therefore directly study their interior structure and the material that surrounds them, thus testing the predictions from numerical simulations.

Our work ties in with a growing list of studies of cosmic density fluctuations based on the comparison of large-scale galaxy and lensing maps \citep[e.g.][]{Planck2013, Waerbeke2013}.  Whereas such studies exploit the statistical correlation between baryonic tracers and lensing fields, we aim for the lensing signal of a spectroscopically preselected sample of voids, rendering our approach equivalent to stacked cluster lensing. While stacking CMB temperature maps at void locations has already been utilized to detect the imprint of the integrated Sachs-Wolfe effect \citep{Granett08.1}, it is the first time that lensing measurements seek to constrain the matter distribution within voids.

We start with a review of the lensing formalism and void model. In \autoref{sec:data} we describe the void and weak-lensing catalogs, in \autoref{sec:results} our novel stacking approach, the resulting void lensing signal, and its significance. We discuss our findings, point out the limitations of our approach, and conclude in \autoref{sec:discussion}.

\subsection*{Approach}

Voids are underdense regions in the matter density distribution $\rho$. 
Gravitational lensing probes this matter field
along some range $\mathcal{D}_z$ in redshift,
\begin{equation}
\Sigma(\theta) = \int_{\mathcal{D}_z} \mathrm{d}z\ \rho(\theta, z)-\bar{\rho},
\end{equation}
being sensitive only to deviations from the cosmic mean density $\bar{\rho}$ -- may they be overdense like galaxy clusters or underdense like voids.
As we will deal with spherical voids\footnote{A brief assessment of this assumption is given in \autoref{sec:discussion}.}, we adopt a radial coordinate frame, in which $\theta=0$ specifies the spatial location of the void center.
In the weak-lensing limit of small perturbations of the matter
distribution, which is certainly justified for voids, the shear
\begin{equation}
\gamma(\theta) =\frac{\Delta\Sigma}{\Sigma_c}=\frac{\bar{\Sigma}(<\theta)-\Sigma(\theta)}{\Sigma_c}
\end{equation}
traces the deviation of the projected surface mass density $\Sigma$ from the average surface density $\bar{\Sigma}(<\theta)$ of all matter inside of a cylinder of radius $\theta$. The critical density $\Sigma_c$
is a function of $\mathcal{D}_z$, the redshift distribution of the
lensed background galaxies, and the angular-diameter distances between
lenses and background. By measuring the gravitational lensing effects of
voids we are therefore able to directly constrain the matter field, but
only in projection along the line of sight. Derivations of the previous equations and details on weak gravitational lensing can be found in e.g. \citet{Bartelmann01.1}.

In this work we will compare our shear measurements to void shapes found in cosmological
simulations. \citet[later called LW12]{LavauxGuilhem2011} determined an
average radial profile,
\begin{equation}
\label{eq:LW12}
\rho(r | R_v) \approx \bar{\rho}\Big(0.13 + 0.70\bigl(\frac{r}{R_v}\bigr)^3\Big),
\end{equation}
in simulations comprising only dark matter particles. The void radius
$R_v$ sets the characteristic scale and is the only free parameter in
this self-similar model. The parameters of the LW12 model are determined for voids of $R_v\simeq8$~\hmpc, but found to describe larger voids similarly well (LW12). By extending this model with a compensation
region outside of $R_v$, \citet{Krause2013} calculated the observable
shear profile for several spherical void models, including LW12\footnote{In fact, \citet{Krause2013} based their calculations on the preprint version of \citet{LavauxGuilhem2011}, which found somewhat shallower void profiles. We recompute the lensing signal analogously to \citet{Krause2013}, but with the LW12 profile as given in \autoref{eq:LW12}.}. We will
use this as the baseline for our comparison to observed shear profiles
around voids in the SDSS footprint.
 
\section{Data sets}
\label{sec:data}

\begin{figure}
\includegraphics[width=\linewidth]{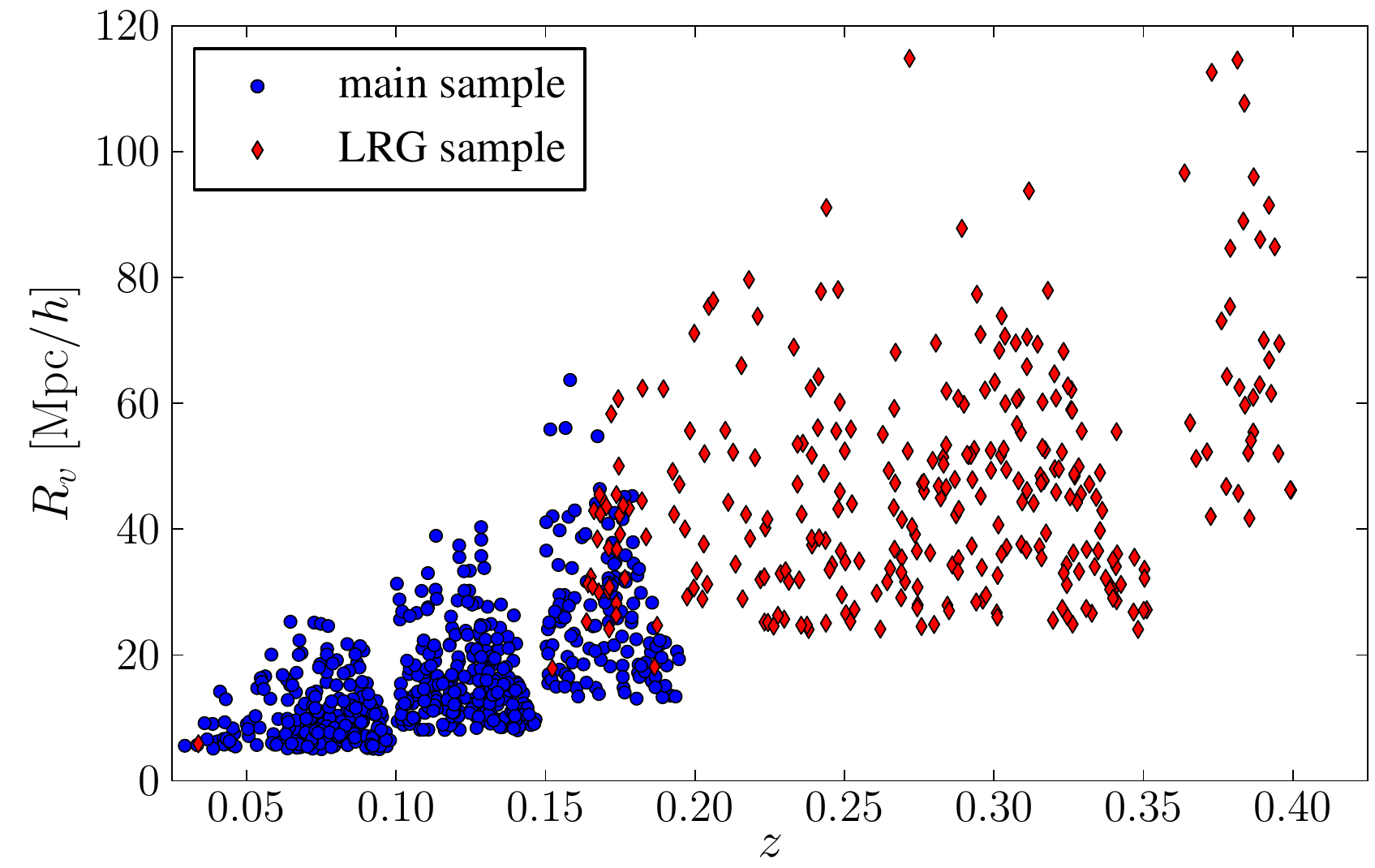}
\caption{Redshift and size distribution of the voids used in this study. For each void, the redshift is taken to be the void center's and the radius is the one of a sphere with the same volume as the void.}
\label{fig:size_redshift}
\end{figure}

We use voids from the \emph{2012.11.17} release of the public cosmic void catalog of \citet{Sutter2012a}\footnote{\url{http://www.cosmicvoids.net}}, based on the ZOBOV algorithm \citep{Neyrinck08.1}. This catalog identified voids in the SDSS DR7 main sample~\citep{Blanton2005} and luminous red galaxy \citep[LRG, from][]{Kazin2010} redshift catalogs, spanning a redshift range of $z=0.0$--$0.45$ and yielding void sizes from $5$ to $120$~\hmpc. To avoid systematics induced by truncated profiles of voids near the survey edges and masks, we take the ``central'' sample, which corresponds to 1031 voids. For 901 of these voids, we can follow the radial shear profiles up to a maximum distance of at least $70$~\hmpc, unobstructed by the survey edges. Their size and redshift distribution is shown in \autoref{fig:size_redshift}.

The weak-lensing measurements are based on SDSS DR8 imaging \citep{dr8}.
Only the $r$-band was used for this study.  The
shear catalog is very similar to that used in \cite{Sheldon09a} and is
an implementation of the ``re-gaussianization'' method from \cite{HirataCalib03}.
We have verified that the shears in this new catalog are consistent with those used
in \citet{Sheldon09a} for objects found in both catalogs.  

The re-gaussianization method
performs well for high S/N galaxies, with expected calibration errors at
less than a percent \citep{Reyes12}, which we have confirmed with a set of simplified simulations.  However, we find that the shears can have ``noise bias'' \citep[e.g.][]{Melchior2012} of several percent for low-S/N galaxies \citep[S/N $\lesssim$ 20, cf. also][]{Reyes12}. Such a calibration bias is multiplicative, and is not expected to be a function of scale when averaged over the full SDSS survey area. Thus this type of bias would not affect the shape of the void signal, but would result in a misestimation of the density in the voids.
To alleviate this effect somewhat, we have
used a more conservative magnitude cut ($r < 21.5$) for this new catalog.
Galaxies at $r=21.5$ have a median S/N of about 10, for which we expect a few
percent calibration error.  Each galaxy receives a weight in the
final analysis $\sim 1/(0.32^2 + \sigma_{e}^2)$, where $\sigma_e$ is the error
in the measured shape and $0.32^2$ is the variance in intrinsic shapes; 
thus, these galaxies get relatively small weight.  Yet they are numerous, so 
there is certainly some remaining calibration bias in the catalog at the few percent level.  However, this error is small compared to the Poisson noise in the void lensing measurements, as we will show below.

\begin{figure*}
\includegraphics[width=\linewidth]{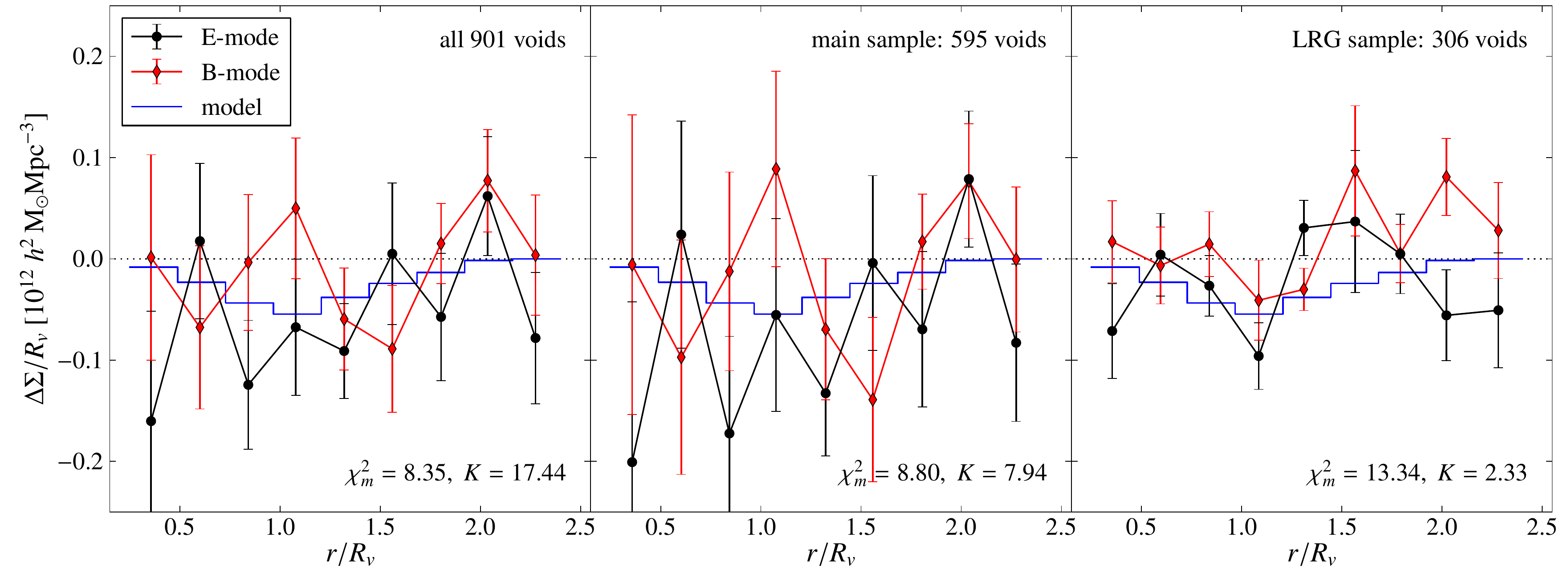}
\caption{Void lensing signal in the range of $r/R_v=0.25\ ..\ 2.4$ of all voids ({\it left}), the main void sample ({\it center}), and the LRG sample ({\it right}). The errorbars show 68\% confidence intervals, estimated from 5,000 bootstrap realizations of the mean in each bin. The blue curve shows the reference LW12 void model calculated from \autoref{eq:LW12}, binned in the same way as the data.}
\label{fig:void_rebinned}
\end{figure*}

We also apply a small ``de-trend'' to the ellipticities in the catalog, subtracting
the mean ellipticity as a function of location in the CCD array and resolution factor $R$, which characterizes the size of the galaxy compared to the PSF width \citep[see][their eq. 8]{Hirata04.1}.
For the bulk of objects, this mean ellipticity is negligible. However, for galaxies with
$R$ values near $\tfrac{1}{3}$, and also for those falling on certain areas of the focal plane for
a subset of our data, the mean ellipticity can be a few percent (\cite{Huff11} for details).
We find that, after de-trending, the catalog
successfully passes various null tests, such as B-mode tests 
and mean shear around random 
points, even on the scales probed in the void analysis carried out here.
Results for B-mode test and random points are shown in \autoref{sec:appendix}.
We also trim the catalog to the 
footprint of the Baryon Oscillation Spectroscopic Survey 
\citep[][BOSS]{BOSS13} and remove galaxies near bright stars 
and in bad fields (where the processing was unsuccessful).

Finally, only those galaxies with a good photometric redshift estimate are
used in the shear analysis (see \cite{Sheldon12} for a description of the catalog selection).
We use the photometric redshift distributions from \citet{Sheldon12} and expect calibration errors associated with these $p(z)$ 
to be less than one percent for the redshifts of interest.  We think
that analysis is sufficient
to characterize the errors for the current work, but note a more realistic treatment of these
errors will be needed for more precise studies.

\section{Results}
\label{sec:results}

As our void catalog comprises voids with size from $5$ to about
120~\hmpc, stacking the measured shear profiles as a
function of the physical distances $r$ would wash out the lensing signal considerably. We therefore rebin the measurements in units of the void radius, $r/R_v$, to take full advantage of the entire void sample, despite the large variations in $R_v$ in the entire sample or any reasonably sized subsample.

In \autoref{fig:void_rebinned} we show the inferred radial surface density contrast rescaled by the void radius, $\Delta\Sigma/R_v$ as a function of the normalized radius $r/R_v$ together with the analytical LW12 void model.\footnote{As this model is self-similar, $\Delta\Sigma/R_v$ is a unique function of $r/R_v$, which means the void model has zero degrees of freedom in our stacking approach. Hence, the model curve in \autoref{fig:void_rebinned} is not a fit to the data.} For the whole sample, the E-mode shows a depression in three consecutive bins around the void. The depression fades towards smaller and larger radii, consistent with the model, but with decreasing significance. The expected depression is also present in the main void sample, albeit at  even larger statistical errors. The LRG sample also shows the depression leading towards $r=R_v$, but returns to a null signal quickly thereafter.

To assess the significance of a lensing detection, we compute the likelihood ratio
\begin{equation}
\label{eq:K}
K \equiv \frac{\mathcal{L}_m}{\mathcal{L}_0} = \exp{\bigl[-\frac{1}{2} (\chi_m^2 - \chi^2_0)\bigr]},
\end{equation}
where $\mathcal{L}_{m/0}$ is the probability of the data being described
by the void model or by the null hypothesis, respectively, and $\chi^2$ is the usual residual sum-of-squares. In this case the null hypothesis can be interpreted literally as signal that is consistent with zero at all scales. Because neither void nor null model have any free parameters to fit, the likelihood ratio is equivalent to the Bayes factor with fair priors, so that our approach is a Bayesian model comparison between two scenarios: One, in which the locations on the sky specified by the void catalog correspond to actual void centers; and the alternative, in which the given locations are entirely random. 

To determine the likelihoods of either hypothesis, we have to make -- and ideally verify -- additional assumptions about the signal and the noise distributions. First,  we adopt a Gaussian for the functional form of the likelihoods, which we found to be a good approximation to the distribution of bootstraps in all bins, although the distributions have substantial power in the tails.

Secondly, as we look for a signal with a specific shape across several bins, it is crucial to determine the amount of covariance between bins. We claim that our measurement is mostly affected by the shape noise of galaxies, for which no bin-to-bin covariance for any single void can exist, as no lensed background galaxy can be counted at more than one radius from the void center. But as we stack many voids, a single galaxy can potentially contribute to more than one void profile, so that a small bin-to-bin covariance in the stacked profile is expected. As long as the voids in our sample are sparse and do not strongly cluster, this correlation will be suppressed due to the randomness of the void center coordinates, so that multiple inclusion of the same galaxy shape would contribute to the stacked profile with effectively randomized orientations. 
Beyond that, the signal is also contaminated with lensing caused by other large-scale structure along the line of sight. While this contribution was properly accounted for in \citet{Krause2013}, performing the bootstraps per void (as opposed to within bins as shown in  \autoref{fig:void_rebinned}), we have verified numerically that the off-diagonal entries of the covariance matrix of the lensing measurements are too small to affect the analysis presented here. These off-diagonal elements are also very noisy in the bootstrap resamples, so that we restrict ourselves to the fairly robust diagonal entries. To account for the fact that the inverse of an thus estimated covariance matrix (even when only diagonal terms are considered) is not an unbiased estimate of the inverse covariance matrix, we corrected the $\chi^2$ values following \citet{Hartlap07.1},
\begin{equation}
\chi^2 \rightarrow \chi^2 \frac{N-B-2}{N-1},
\end{equation}
where $N$ denotes the number of voids in the sample and B denotes the number of bins of the profile.

Thirdly, one has to bear in mind that the void model, and hence $\chi^2_m$, additionally assumes that the void radii are accurate and that the void lensing signal is characterized by the LW12 profile. We will discuss the implications for our analysis in \autoref{sec:discussion}.

For the full sample, we find $K=17.44$ and a model error of $\chi^2_m=8.35$. With nine independent bins, the model constitutes an excellent fit to the data and is clearly preferred over the null.  Due to substantially larger errors, the main sample is less decisive with $K=7.94$. As mentioned above, there is only weak evidence of lensing in the LRG sample, reflected in a rather poor $\chi^2_m=13.34$ and a likelihood ratio $K=2.33$.

As a measure of potential systematic contamination, we also show the B-mode in \autoref{fig:void_rebinned}. While it is not zero at all scales, its largest deviations for the full sample occur on the small-scale side where the errors are largest due to a small number of background galaxies. Overall, we find $\chi^2_0 (B) = 6.83$, clearly consistent with a null signal. As the B-mode fluctuations appear uncorrelated with the E-mode, we do not expect them to drive  the lensing signal. 

Taken at face value, we have substantial evidence of lensing in the sample of all voids, and weaker evidence in the main sample. But there are some aspects worthwhile mentioning about the validity of this result. It seems odd that the whole sample has much larger $K$ without having much smaller $\chi^2_m$ compared to the main sample, but this is in fact characteristic of the likelihood ratio, whose power to reject the null grows exponentially with sample size if the alternative is true. 
It is also counter-intuitive that the errorbars for the LRG sample are substantially smaller than for the main sample, so that the larger voids appear in principle better suited to pick up the lensing signal. Because the noise in this measurement primarily stems from the shape scatter of the background galaxies, one would assume that voids at lower redshift with more galaxies behind them should have smaller errors. There are also more voids in the main sample than in the LRG sample. But the profiles in \autoref{fig:void_rebinned} are plotted as $\Delta\Sigma/R_v$, which also rescales the per-void errors. In addition, the distances $r$ are also rescaled so that radial bins correspond the annuli on the sky, whose area is proportional to $R_v^2$. For the lensing data at hand, these two effects more than make up for the lower number of voids and background galaxies of the LRG sample. We will point out in \autoref{sec:discussion} why this is still not enough to allow for a clear measurement of the lensing signal.

\begin{figure}
\includegraphics[width=\linewidth]{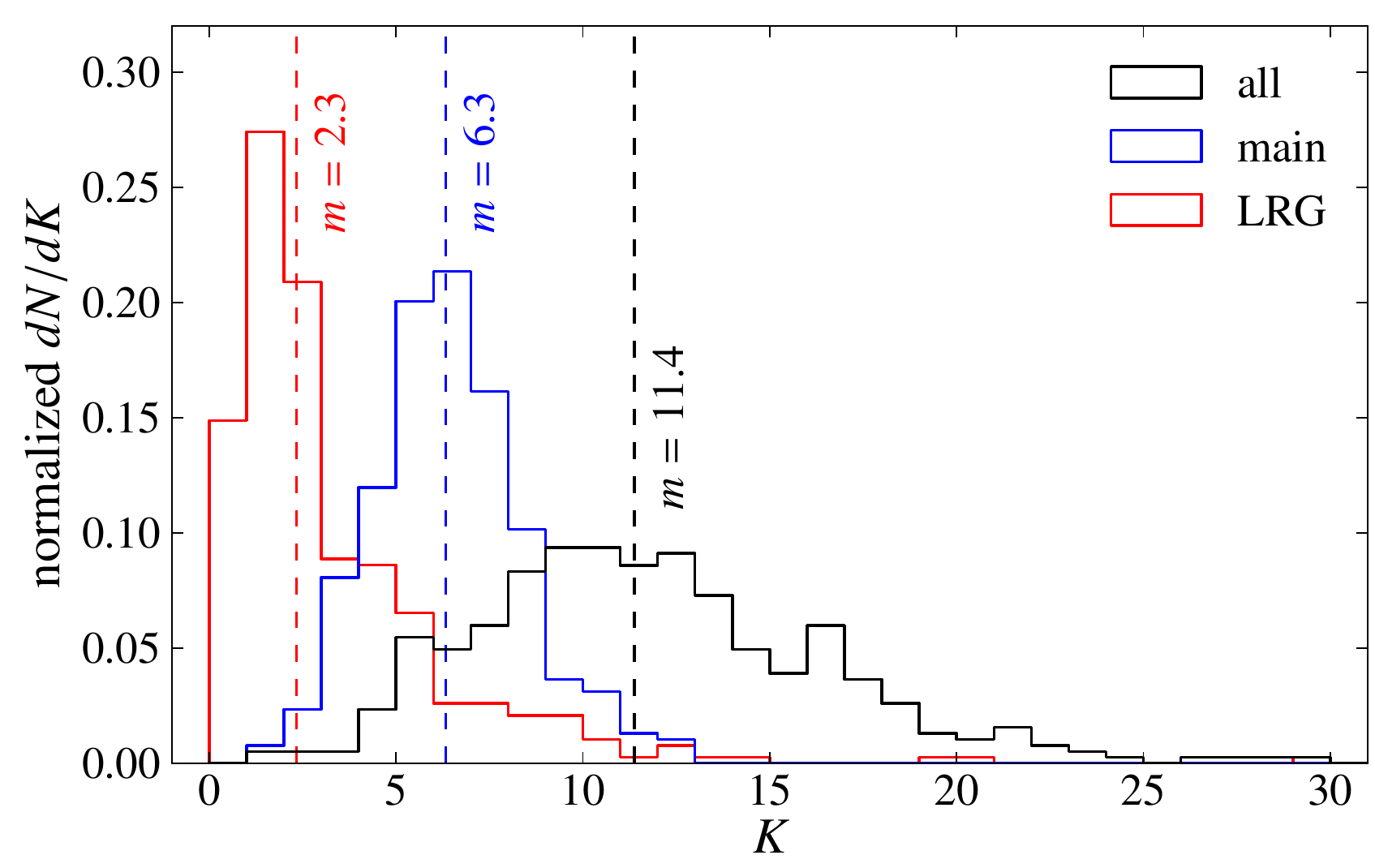}
\caption{Distribution of likelihood ratios $K$ in the entire sample (black) and the main (blue) and LRG (red) samples. The spread is caused by varying the binning in units of $r/R_v$, not from the bootstrap resampling, which results in $\Delta K\approx 0.1$ and is therefore negligible here. The vertical dashed lines and the numbers indicate the median $m$ of either distribution.}
\label{fig:K_hist}
\end{figure}

Finally, the most problematic aspect of the analysis lies in the choice of the binning in \autoref{fig:void_rebinned}. Precisely because we deal with a weak signal compared to the noise, it matters a lot in which bins the lensing measurements happen to fall. Given a particular binning, such as the one shown above,  statistical fluctuations may or may not cancel each other to yield a good or a poor estimate of the mean in any of the bins. So we need to expect a substantial variation of $\mathcal{L}_{m/0}$ and consequently $K$. Indeed, when varying the binning scheme within plausible limits of $\min(r/R_v)=\lbrace 0.0, 0.1, ..., 0.5\rbrace$ and $\max(r/R_v)=\lbrace1.6, 1.7, ..., 2.5\rbrace$ using a number of bins $B+1=\lbrace6,8,...,16\rbrace$,  \autoref{fig:K_hist} shows a very broad distribution of $K$, in particular for the whole sample. It does, however, qualitatively support the findings we made earlier: The whole sample has substantial evidence for lensing with a median $K=11.4$, while the main sample has weaker evidence with a median $K=6.3$. None of these two distribution has considerable power between $K=0 .. 1$, so that we can conclude that with the data we have at hand, the evidence for lensing is not simply a fluke based on a lucky choice of the binning scheme. The LRG sample hardly goes beyond $K=3$ and even has a 15\% probability of $K<1$, which would favor the null hypothesis.

\section{Discussion and conclusion}
\label{sec:discussion}

While we see a substantial depression of the density contrast around the void radius, there are several potential limitations to our analysis, all of which concerns different aspects of the void model being the correct description of actually observed voids.

First, the assumption of self-similarity. It allowed us to rebin the lensing data in 
terms of $\Delta\Sigma/R_v$ and $r/R_v$ and thus stack all voids, 
irrespective of their actual size, on top of each other. Given the 
small size of the void sample and the low number density of background 
sources in the lensing data, this rebinning turned out to be crucial 
for a significant lensing detection. 

If, on the other hand, voids are not self-similar, the approach we took would mix voids with different profiles and hence reduce the statistical power of the test. The same happens when the estimates of the void radius are inaccurate or the stack of voids is not perfectly spherical due to sample variance. Taken to the extreme, our assumed model could be such an inaccurate description of the measured stacked lensing signal that the likelihood ratio $K$ would be in favor of the null, even if the void catalog provides valid void locations and these voids properly act as (anti-)lenses. This is not the case here. While we do not claim that voids are necessarily self-similar in nature or that the void radii are precisely estimated or that there is no residual deviation from average sphericity, these conditions seem to be fulfilled well enough to enable our approach.

Secondly, in addition to self-similarity we have adopted a particular, perfectly compensated void model, which means that when integrating out to a sufficiently large radius, the enclosed matter will have exactly average cosmic density. Specifically, \citet{Krause2013} assumed a constant density compensation region between $1.0921R_v$ and $2\,R_v$ such that the void is compensated at this outer radius. This has implications for the shape of the expected shear profiles, mainly for the value of the profile around $r=R_v$ (see their Fig. 3 for details). 
In fact, recent studies of the galaxy distribution in SDSS and in realistic mock simulations have revealed that small voids (especially with $R_v\leq10$~\hmpc) tend to be overcompensated, whereas large voids have very little compensation at all \citep{Ceccarelli2013, Hamaus2013}.
Given the limited significance of our results, we can neither support nor reject such a dichotomy on the basis of a lensing measurement, which would be free of galaxy bias.

Finally, we address the lack of a significant lensing signal in the 
LRG sample. Because large voids should not be fully compensated, they should generate an even more negative lensing signal than predicted
by our compensated model, also extending far beyond $2 R_v$, which would make it easier to pick up than in the main sample. However, the sparse sampling of the galaxy field by LRGs introduces additional uncertainties for the void finding algorithm. In a dedicated simulation study, we found that optically detected voids in the LRG sample still correspond to underdensities in 
the dark matter distribution, but the average level of underdensity is reduced by about a factor 2, at least within $\tfrac{1}{2}R_v$ \citep{Sutter14.1}. As a consequence, the lensing signal would suffer a similar degradation. Due to the sparse sampling, we also expect larger uncertainties in the void radius estimate, which would reduce the significance of the data in our rebinning approach even further. These two issues can make LRG voids suboptimal for a lensing detection and hence render their lensing signal undetectable given the limited statistical power of our data set. 

In summary, by using a spectroscopically selected void sample and a well-tested
shear catalog together with a novel rebinning technique, we were able to detect
the lensing signal arising from the underdensities of cosmic voids in the SDSS DR7 
footprint with a median likelihood ratio of 11.4:1 over a random null signal. Even with the largest currently available data sets for this kind of analysis, it remains a rather weak detection. Improvements to our analysis require larger void samples or substantially deeper lensing surveys, both of which can be achieved in the upcoming years. The practical difficulty stems from having to do void finding and weak lensing in the same footprint.
 
We believe our findings to be robust despite the overall low significance of the stacked lensing signal, and invite independent analysis by releasing the data we have used in this work together with the stacking and bootstrapping code here: \href{https://github.com/pmelchior/void-lensing}{https://github.com/pmelchior/void-lensing}.

{\small
\section*{Acknowledgements}
The authors thank Guilhem Lavaux and Christopher Bonnet for useful discussions. The manuscript profitted considerably from suggestions by Uro{\v s} Seljak and the anonymous referee. 
PM is supported by the U.S. Department of Energy under Contract No. DE- FG02-91ER40690.
PMS and BDW acknowledge
support from NSF Grant AST-0908902. BDW acknowledges funding from an 
ANR Chaire d’Excellence, the UPMC Chaire Internationale in Theoretical 
Cosmology, and NSF grants AST-0908902 and AST-0708849.
ES was supported in part by the U.S. Department of Energy under
Contract No. DE-AC02-98CH10886.}

{\small
\bibliography{lensing}
}

\appendix

\section{B-mode and random point signal}
\label{sec:appendix}

\begin{figure*}
\includegraphics[scale=0.5]{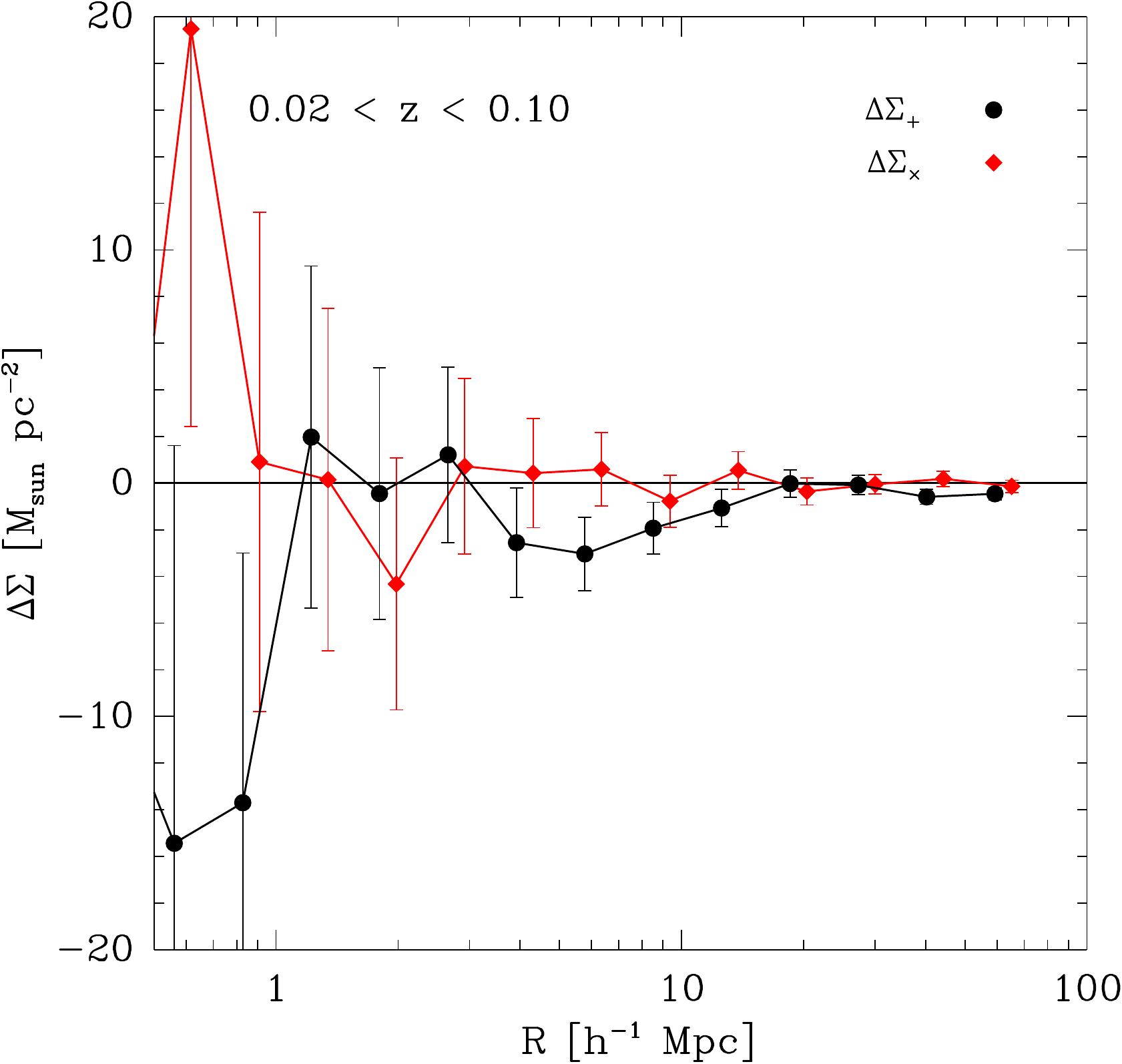}
\includegraphics[scale=0.5]{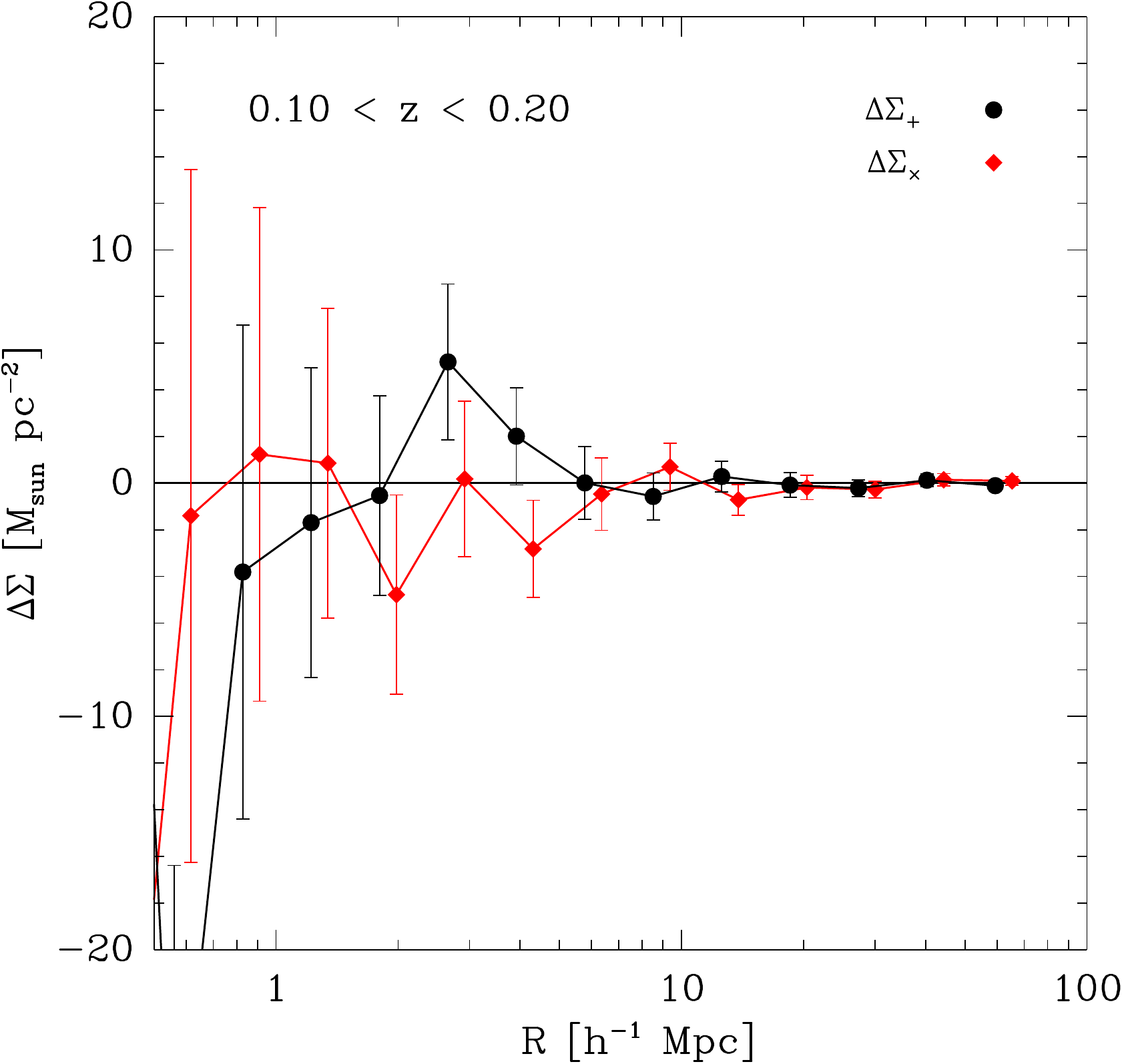}
\caption{Lensing E-mode ($\Delta\Sigma_+$, \emph{black}) and B-mode ($\Delta\Sigma_\times$, \emph{red}) for voids in two separate redshift bins.  The signal was averaged as a function of physical radius rather than in units the void radius.}
\label{fig:void_bmode}
\end{figure*}

\begin{figure*}
\includegraphics[scale=0.5]{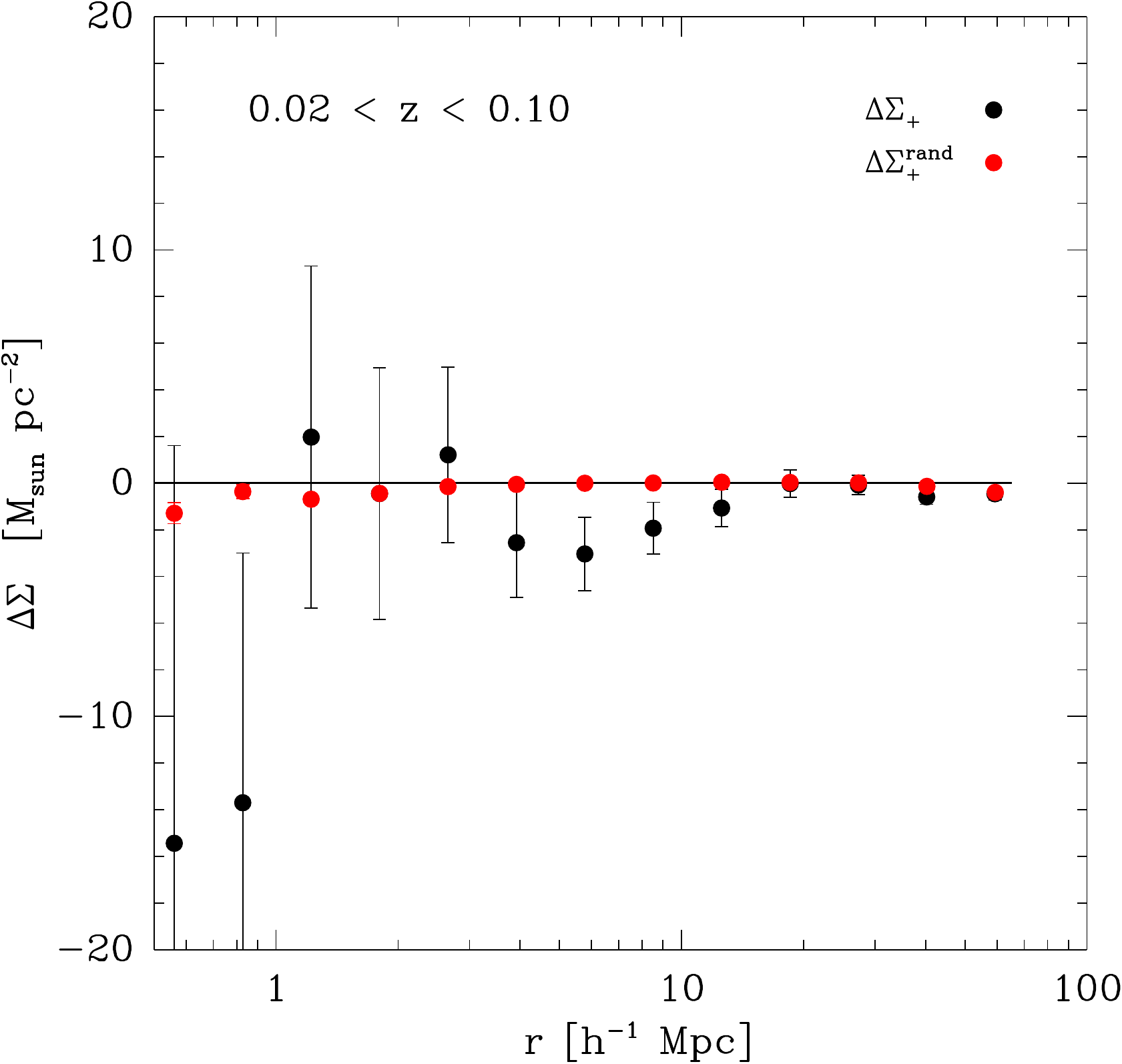}
\includegraphics[scale=0.5]{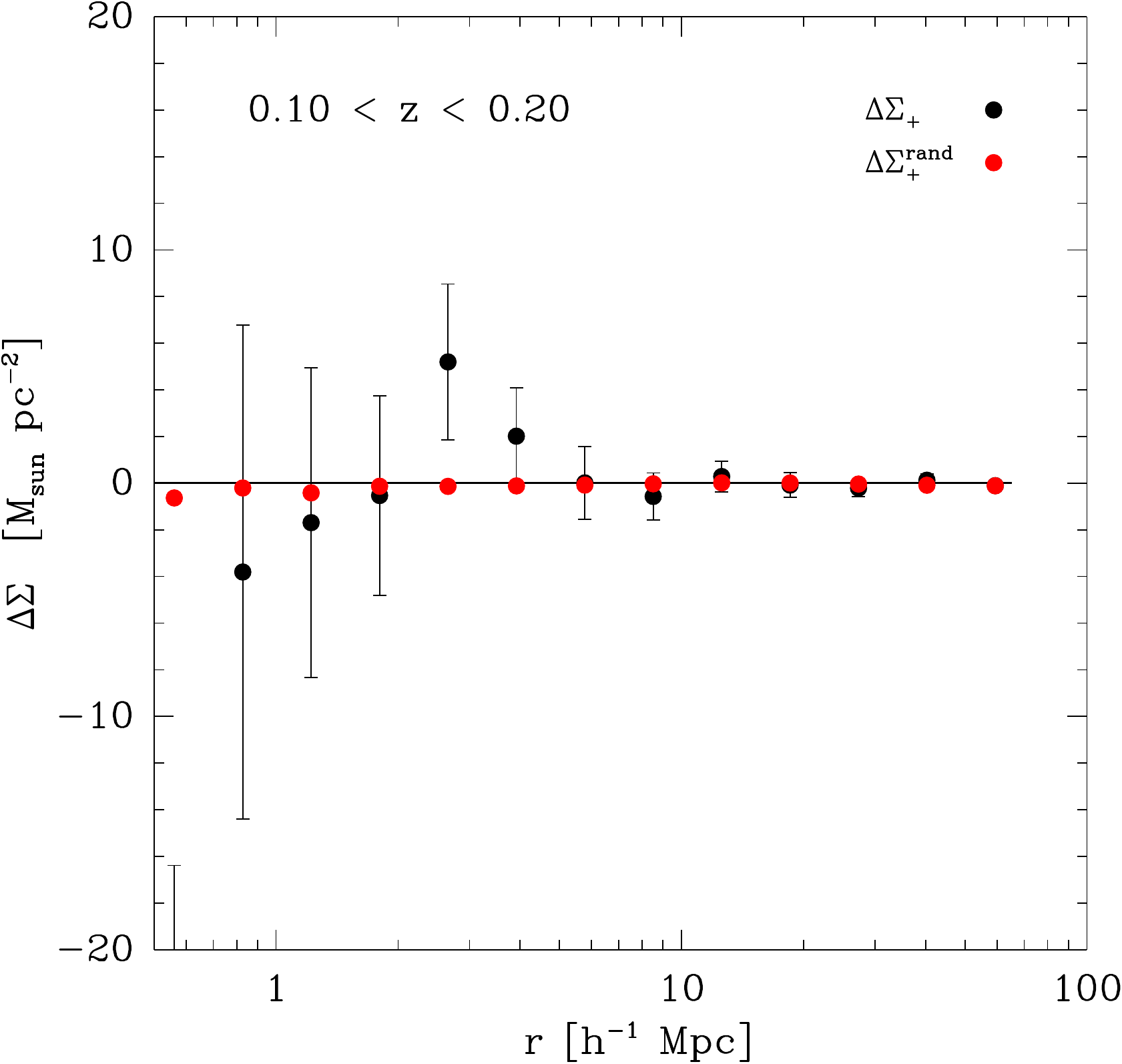}
\caption{Lensing signal around voids (\emph{black}) to the signal around random points (\emph{red}) for  two different redshift ranges. The signal was averaged as a function of physical radius rather than in units the void radius. }
\label{fig:void_rand}
\end{figure*}

In \autoref{fig:void_bmode} we compare the B-mode
signal (labeled with $\Delta\Sigma_\times$) to the E-mode signal (labeled $\Delta\Sigma_+$).
We expect no B-mode if the signal is created by gravitational lensing.  The stacked signal is shown for voids drawn from two separate redshift ranges $[0.02,0.10]$ and $[0.10,0.20]$.
In contrast to \autoref{fig:void_rebinned}, the radial binning is in physical units rather than units of the void radius. We find no significant detection of B-modes in either redshift bin, consistent with our findings in \autoref{sec:results}.

In \autoref{fig:void_rand} we compare the signal the E-mode signal measured around voids to that measured around random points in the survey footprint. The footprint is taken directly from the SDSS DR7 public release and is the same used to identify boundaries when finding voids (see Fig. 3 of \citealt{Sutter2012a}). Redshifts were drawn uniformly in the volume from redshift zero to redshift 0.3 and weighted to match the redshift histogram of the voids.  There is a detection of a small signal around random points at small and large scales, but the amplitude is not large enough to account for the detected void signal. We subtract the mean random points signal from the signal around voids in the analysis presented above.

\label{lastpage}
\end{document}